# Democratising Risk:

# In Search of a Methodology to Study Existential Risk

Carla Zoe Cremer[i]  Luke Kemp[ii]

**Abstract**

Studying potential global catastrophes is vital. The high stakes of existential risk studies (ERS) necessitate serious scrutiny and self-reflection. We argue that existing approaches to studying existential risk are not yet fit for purpose, and perhaps even run the risk of increasing harm. We highlight general challenges in ERS: accommodating value pluralism, crafting precise definitions, developing comprehensive tools for risk assessment, dealing with uncertainty, and accounting for the dangers associated with taking exceptional actions to mitigate or prevent catastrophes. The most influential framework for ERS, the "techno-utopian approach" (TUA), struggles with these issues and has a unique set of additional problems: it unnecessarily combines the study of longtermism and longtermist ethics with the study of extinction, relies on a non-representative moral worldview, uses ambiguous and inadequate definitions, fails to incorporate insights from risk assessment in relevant fields, chooses arbitrary categorisations of risk, and advocates for dangerous mitigation strategies. Its moral and empirical assumptions might be particularly vulnerable to securitisation and misuse. We suggest several key improvements: separating the study of extinction ethics (ethical implications of extinction) and existential ethics (the ethical implications of different societal forms), from the analysis of human extinction and global catastrophe; drawing on the latest developments in risk assessment literature; diversifying the field, and; democratising its policy recommendations.

[i] The Future of Humanity Institute (FHI), University of Oxford.
[ii] Centre for the Study of Existential Risk (CSER), University of Cambridge.
*Equal authorship. The ordering has been arranged alphabetically.*

# 1. Introduction

Over the past two decades, scholars have begun to methodically study human extinction and global catastrophes. This field of "Existential Risk Studies" (ERS) aims (i) to identify existential and catastrophic risks; (ii) map out the potential causes of existential catastrophes; (iii) understand the ethical implications of such calamities and (iv) devise effective strategies for mitigation and prevention.

Although the field is relatively small, it has expanded considerably, especially over the past ten years[1]. It is also of increasing public interest: several popular trade-books have been published[2–7]. The ideas have been integrated into vision-setting reports from the UN Secretary General[8]. Institutions focusing on existential risk have received hundreds of millions of dollars in philanthropic funding[iii]. It is a field on the rise.

It is commendable and overdue that the study of human extinction is receiving greater academic engagement and public attention. However, this field needs to be held to high standards: it is ambitious, could affect the lives of many and attracts scholars who seek to change the trajectory of global society.

The field faces daunting challenges. How can it be inclusive of the diversity of human preferences and visions of the future? How can researchers avoid baking their subjective assumptions into risk analyses that might affect those who do not share their values? How do they conduct complex risk assessments? How do they deal with uncertainty? How do they compare risks with different quantities of evidence and degrees of plausibility? How do they ensure that the catastrophes the field studies are not misused to justify dangerous actions?

The field has not yet established the answers to such questions and we are not the first to be aware of this[9]. Throughout this paper we will point to the scholars who we know have raised similar questions. The historically dominant techno-utopian approach (henceforth the "TUA") played an important role in establishing the field and drawing attention to the significance of studying human extinction. It is time to examine this approach with a critical eye. We do this to identify weaknesses, areas for further investigation and the need to also explore alternative approaches. The TUA, which relies heavily on total utilitarianism, transhumanism, and (strong) longtermism, is too morally unrepresentative, methodologically

---

[iii] The following is not a comprehensive summary, but an indicator of the origin of resources in the field: The Future of Life Institute began with a $10 million donation from billionaire Elon Musk. It launched a $25 million program on AI safety with financial support from billionaire Vitalik Buterin. The technology millionaire Jaan Tallinn was a co-founder of CSER, while Elon Musk sits on the Advisory Board. The Future of Humanity Institute has received funding from Elon Musk, as well as over $14 million in grants from Open Philanthropy. Open Philanthropy, founded by billionaires Cari Tuna and Dustin Moskovitz (a cofounder of Facebook), funded the establishment of the Centre for Security and Emerging Technology (CSET) with a $55 million dollar donation. Open Philanthropy has, to date, given over $37 million to projects related to global catastrophic risks, $99 million to biosecurity and pandemic preparedness, and $196 million worth of grants for potential risks from AI (not all of which are related to existential risk) (calculated on the 01/12/2021 using Open Philanthropies online database).

flawed, and risky an approach to rely on. It is time to diversify the definitions, tools, and frameworks of ERS. We need to develop an ERS methodology that addresses the core questions of ERS and avoids the problems of the TUA.

The question we raise is: what should the study of human futures and catastrophe look like under moral uncertainty? We suggest some solutions: a diverse range of approaches, deliberative democratic processes, and the separation of the study of catastrophe and extinction from the ethics of human existence and extinction.

We proceed in Section 2 by outlining the moral assumptions of the TUA. In Section 3 we give reasons why the TUA is not representative of wider human preferences and why representation matters to the goals of ERS. Section 4 explains how moral and empirical assumptions are often masked by abstract and ambiguous definitions and tools. Moral and empirical assumptions are important because they can distort the results of our work, whether it be in how we conceive of risks, which risks we prioritise, or what policies we recommend. In Section 5 we focus on how studying existential risk could backfire by growing other sources of catastrophic risk. In Section 6 we suggest a democratic, risk-averse approach.

## 2. The Techno-Utopian Approach to ERS

We focus on the techno-utopian approach to existential risk for three reasons. First, it serves as an example of how moral values are embedded in the analysis of risks. Second, a critical perspective towards the techno-utopian approach allows us to trace how this meshing of moral values and scientific analysis in ERS can lead to conclusions, which, from a different perspective, look like they in fact increase catastrophic risk. Third, it is the original and by far most influential approach within the field.

### 2.1 Definitions and History

#### 2.1.1 The Influence of the TUA

The TUA is a cluster of ideas which make up the original paradigm within which the field of ERS was founded. We understand it to be primarily based on three main pillars of belief: transhumanism, total utilitarianism and strong longtermism. More precisely: (1) the belief that a maximally technologically developed future could contain (and is defined in terms of) enormous quantities of utilitarian intrinsic value[6,10–13], particularly due to more fulfilling posthuman modes of living[14,15]; (2) the failure to fully realise or have capacity to realise this potential value would constitute an existential catastrophe[6,11,13]; and, (3) we have an overwhelming moral obligation to ensure that such value is realised by avoiding an existential catastrophe[16], including through exceptional actions[11,17].

Not all publications that make use of the TUA explicitly support every element of the approach, but the most widely read publications incorporate a significant number of TUA elements and share its visions of the long-term future.[iv] The most popular definitions of existential risk are still the initial techno-utopian definitions by Bostrom and more abstract, but very similar versions based on expected value (see Section 4.1). The few attempts to put forward alternative frameworks[18,19] are not nearly as widely cited, known or used[v]. Importantly, they do not offer alternative definitions of what an existential risk is. Despite the increase in size and diversity in the field of ERS there appears to still be no coherent alternative to the TUA.

The impact of the ideas of the TUA can be seen across the field, characterising its most cited and best-known publications. Beard and Torres trace the beginning of the existing field of ERS to the early publications by Bostrom[21]. It was Bostrom's work in the early 2000s that first aimed to formalise the concept of an existential risk, notably via his 2002 paper "*Existential Risks: Analyzing Human Extinction Scenarios and Related Hazards*"[13] and the 2013 article "*Existential Risk Prevention as Global Priority*"[11], which have been cited a combined 1,046 times[vi]. These two papers articulate the canonical definitions of "existential risk" and, along with Bostrom's 2003 paper "*Astronomical Waste*"[12], present the clearest distillation of the TUA.

The TUA also characterises almost every existential risk text with significant public profile. This includes trade-books such as *Superintelligence*[3], *The Precipice*[6], *Life 3.0* [4], and *What We Owe the Future*[22]. Several culturally influential ideas such as a technological singularity[23,24] and longtermism[25] are influenced by the TUA. The techno-utopian worldview also appears to resonate with key funders. For example, Holden Karnofsky, the co-founder and co-chief executive of Open Philanthropy, strongly echoes the TUA in his "Most important century" series[26].

The TUA we describe should be understood an ideal type[vii]: both the texts and thinkers under it may vary in specifics but converge in their broad vision. This is not an issue: despite many national variations we can speak of capitalism in general and recognise particular countries as being capitalist.[27]

### 2.1.2 Defining Existential Risk under the TUA

Bostrom provides two general formulations of existential risk. He initially defined it as "where an adverse outcome would either annihilate Earth-originating intelligent life or permanently

---

[iv] The failure modes we describe in our paper will thus apply to the elements of the TUA which different publications lean on.

[v] Liu et al.[19] (2018) suggest examining GCRs in terms of their hazards (what harms humanity), vulnerabilities (how and why a given hazard could cause humanity to come to harm) and exposure (how and why humanity is in harm's way). Avin et al.[18] classifies GCRs by the critical system under threat, the mechanism for spreading risk globally, and failures to mitigate or prevent the risk. See Beard et al[20] for a brief summary, as well as a survey of existing definitions for GCRs.

[vi] Calculated with Google Scholar on the 10th May, 2021. As an imperfect comparison, one of the few papers proposing an alternative way of thinking about existential risks (Liu et al, 2018[19]) has a total of 24 citations, (calculated on the 10th May, 2021).

[vii] A Platonic ideal of a phenomenon that abstracts its essential features but in reality is observed in many variations.

and drastically curtail its potential"[13]. Later, he provided a more refined definition: "one that threatens the premature extinction of Earth-originating intelligent life or the permanent and drastic destruction of its potential for desirable future development"[11]. The latter paper provides a further typology of existential risks as:

   i. Premature extinction of humankind before reaching technological maturity;

   ii. Failure to reach technological maturity due to an unrecovered collapse, recurrent collapse, or plateauing;

   iii. Technological maturity being realised in an irredeemably flawed manner; or,

   iv. Ruination of Earth-originating intelligent life after technological maturity is reached but before its full potential can be realised.

According to this typology, the core feature of an existential catastrophe is a failure to attain a stable state of technological maturity, maintained for as long as is physically possible. Bostrom specifies this as a level of technological development, resource acquisition, and resource efficiency that allows for the highest feasible level of "economic productivity and control over nature".[11] Failure is anything but the full exploration of possible options and the full exploitation of available matter.

A recent definitional reworking of this approach is the formulation of an existential catastrophe as "an event which causes the loss of a large fraction of expected value"[28] or less technically as a risk that "threatens the destruction of humanity's longterm potential"[6].

### 2.2 Transhumanism: Humans as a Stepping Stone

Transhumanism is the moral position that there is value in exploring posthuman and transhuman modes of being[15]. The results are to be beings — modified biological humans, cyborgs, androids, or digital simulants — whose lives are considered more valuable than current ones[11]. Transhumanists argue that these beings could achieve far longer, richer lives marked by net positive experiences[14].

Achieving such lives would depend on further technological progress.[15,29] We would need three fundamental transformations: (i) protecting life; (ii) expanding cognition, and (iii) elevating well-being[14]. This can, for example, take the form of achieving immortality, superintelligence, and a greater capacity for pleasure.

Although transitioning to a posthuman stage could of course entail the extinction of *Homo sapiens*, Bostrom contends that "the permanent foreclosure of any possibility of this kind of transformative change of human biological nature may itself constitute an existential catastrophe".[11] Preventing existential risk is not primarily about preventing the suffering and termination of existing humans, it is focused on preserving humans so that they may give rise to a post-human species that contains more value.

## 2.3 Total Utilitarianism: Humans as Containers

Total utilitarianism identifies moral rightness with the maximisation of well-being. Well-being could be interpreted in hedonistic, desire-satisfactionist, or objective-list theory terms[30]. People thus carry some unit of value and the greater this value, the better. Total utilitarianism therefore demands that we maximise the total amount of value in the universe, with as many people coming to exist as possible, each person living an overall happy (i.e., net-positive) life, regardless of where or when these people come into existence. This equivalence in moral patient-hood between different "containers" of value here relies on the "impersonalist" or "non-identity" perspective, in which it is not relevant who is affected, only that someone is affected.

The utilitarian argument that the future should be an overwhelming moral priority relies on an assumption that the number of intelligent beings who could come to exist could be unimaginably large. Matheney for example, estimates a low-range figure of $10^{16}$ assuming we remain Earth-bound[10]. Bostrom argues that if our descendants colonised as much of the universe as quickly as possible and converted celestial bodies into "computronium" for running simulations, this could result in some $10^{38}$ simulated conscious beings per century in Earth's supercluster alone[11]. Assuming wider interstellar exploration then we could even produce $10^{58}$ happy simulations.[3] If we are alone in the universe and computations are run at colder temperatures towards the end of the heat death of the universe then even more could be achieved[31].

The posthuman calculus, or the "astronomical value" thesis, is that the future could hold far greater value than the present due to the teeming masses of beings with higher levels of happiness than organic mortals can achieve. *Homo sapiens* is eclipsed by the towering shadow of the techno-utopian future.

## 2.4 Strong Longtermism: The Present in the Shadow of the Future

The third philosophical foundation is strong longtermism. Considering future generations as moral patients is by no means new. Notions of intergenerational justice, equity and fairness across the deep future have been extensively discussed for decades.[32] Strong longtermism goes a step beyond this and suggests that for some situations we may have an ethical imperative to select the choices expected to have the best effect on the long-run future[16], and usually relies on a utilitarian calculus to justify this.

The best choice is often equated with the choice that has the highest expected value[33]. Expected value is calculated by multiplying the value of an outcome by the probability of it occurring. A calculus that numerically favours strong longtermist actions, such as reducing existential risk, rather than saving millions of today's people, often relies on the assumption of continued technological development, happy future people, and interstellar settlements[11,33]. Thus, ensuring technological progress and maximising the quantity and expected quality of hypothetical future lives may be deemed more important than protecting current lives. We have no principled guidance about when and why a strong longermist should prioritise living humans of today.

Transhumanism, total utilitarianism, and strong longtermism are a coherent, re-enforcing and complementary set of beliefs. Some versions of longtermism might be compatible with multiple theories of value[6], but this is an ongoing area of study[34–36]. It is unclear how different those versions would be and what this would practically imply[16]. The reasons put forward for why other moral theories should care about existential risk include to cultivate civilisational virtues and/or to meet intergenerational obligations[6]. These reasons explain why different moral theories should be concerned about human extinction, but not why they would support strong longtermism.

The non-utilitarian case for strong longtermism is, for now, weak. There have been too few attempts to understand whether different moral positions can support the belief that the vast majority of value lies in the future. There is countering evidence that this is not the case[37–39]. While it may be the case that a wide range of moral positions support caring about the future in a broad sense, it is not necessarily the case that each of these views yield the same practical implications. Hence, it is not appropriate to argue that the implications of strong longtermism follow from a diverse range of moral positions.

## 3. Representation and Existential Risk
### 3.1 The TUA as a Non-Representative View

The TUA is not representative of what most humans alive now believe. Relying on the TUA, which is unrepresentative of many people's moral views today, can distort the analysis of existential risk. Representativeness itself says nothing about whether its philosophical pillars are wrong or right, but it is risky to rely exclusively on one unrepresentative approach given moral, political and empirical uncertainty. Theoretical work in ERS should be paired with and constrained by empirical studies that capture the range of existing intuitions about human extinction and longtermism. We must know the moral intuitions of the public, and when experts dismiss their moral intuitions as incorrect, they must have strong arguments to do so.

There is no consensus among philosophers on moral theory. Utilitarianism is not the most commonly held view In one of the few surveys in the area less than a quarter (23.6% ) of philosophers identified with consequentialism[40]. An even smaller number will be utilitarian, and a small number still will be total utilitarian. Techno-utopia offers futures of pleasurable, often virtual experiences, in which commonly valued attributes like purpose, virtue, love, and justice do not play a central role.

Transhumanism, too, is a niche perspective, and surveys reveal that those who identify as transhumanists come from a narrow demographic. The most recent high-quality survey, one that collected 760 responses from members of the World Transhumanist Association (now called Humanity+) in 2007, found that 90% of the respondents were male with a median age of 30-33 years old[41]. It is unknown how many people of the wider population would accept all or some of the premises of transhumanism if they were surveyed.

The implications of the TUA definition of existential risk also appear to be unrepresentative. According to the original definition of existential risk (a failure to attain technological maturity) technological plateauing, or a failure to spread beyond Earth, is an existential risk. Both near-term extinction due to nuclear war and a future in which humans persist sustainability and equitably for the next billion years without major technological progress are seen as existential risks: worst-case outcomes for humanity. Equating these outcomes as morally equivalent is likely unintuitive to many[42].

The perspective that potential future lives are morally equivalent to existing lives may also be unintuitive to many. It is an active area of theoretical debate between philosophers, and we need more surveys that empirically query the moral intuitions of a wider population. Caviola et al.[43] find context-dependent support for adding future people. Schubert et al.[44] find context-specific overlap between surveyed intuitions of lay-persons and theoretical arguments by experts in ERS, although this depends heavily on survey framing. These are a commendable start. However, far more work is needed to understand what the wider population of the world wants from the far future. Ideally, this should be built not just on surveys, but also more deliberative practices (see Section 6).

The techno-utopian vision of the future, which combines three rather uncommon positions (transhumanism, total utilitarianism and strong longtermism) and considers technological stagnation to be an existential risk, is likely a rare view among the global population. It may rise in popularity in the future, but presently it appears to be a fringe position.

### 3.2 The Risk of a Non-Representative View

Tying the study of a topic that fundamentally affects the whole of humanity to a niche belief system championed mainly by an unrepresentative, powerful minority of the world is undemocratic and philosophically tenuous. Landemore defines the term "elite" as a group of people that would not likely be selected at random from its wider population and that is granted decision-making powers[45]. Under this definition, the field of existential risk is decidedly elitist at present. There are ways to mitigate against elitist research projects: diversifying the field and thus its policy recommendations, and democratising the evaluation of policies that are proposed by the field (see Section 6).

An obvious retort here would be that these are scholars, not decision-makers, that any claim of elitism is less relevant if it refers to simple intellectual exploration. This is not the case. Scholars of existential risk, especially those related to the TUA, are rapidly and intentionally growing in influence. To name only one example noted earlier, scholars in the field have already had "existential risks" referenced in a vision-setting report of the UN Secretary General. Toby Ord has been referenced, alongside existential risks, by UK Prime Minister

Boris Johnson. Dedicated think-tanks such as the Centre for Long-Term Resilience have been channelling policy advice from prominent existential risk scholars into the UK government[46].

The field also appears to be elitist in the more common-sense notion of being representative of a small stratum of people with disproportionate economic and political power. The main research centres are clustered in a few of the most elite universities in the world, with most of the field located in Oxford, Cambridge, or the San Francisco Bay Area. As noted earlier, the field is disproportionately supported by billionaires and millionaires. They don't only hold financial power, but often advisory positions as well. The ideas of the TUA also closely echo popular Silicon Valley ideology[47,48]. Indeed, in 1995 Barbrook and Cameron referred to the techno-utopian ideas of the 'Californian Ideology', which distinctly echoes the TUA. The Californian Ideology began in Silicon Valley during the tech boom of the 90s and was underpinned by a commitment to technological determinism and neoliberal economics[49].

The point is a broader one: it is highly risky to grant privileged influence over the fate of *Homo sapiens* to a tiny minority. This is true for the study of existential risk, but more so for the implementation of policies that are meant to reduce existential risk, which will need to balance trade-offs between different interests. An attempt to reduce elitist interference in the study and implementation for existential risk mitigation is important due to moral uncertainty[50]. For every moral theory, there exist recommendations which fail to match our intuitions[6]. It is thus all the more important to know what choices would empirically be preferred by widening the range of people that are allowed to decide what risks are worth and not worth taking.

Some scholars associated with the TUA have written about moral uncertainty. This is excellent and should continue. They advance theories such as expected moral value (ranking alternative axiologies by their expected value[36], or expected moral choice-worthiness (weighting moral theories by credence and combining them)[50] to navigate moral uncertainty. Issues remain for these approaches, including whether moral theories are comparable in this manner, whether empirical and moral uncertainty are equivalent, how such approaches are scaled up to collectives, as well as technical problems[51]. Using the suggested approaches tends to lead to total utilitarian perspectives outweighing others once large populations in the future are assumed.[36] There is room for considerably more research, and approaches to dealing with moral uncertainty have yet to be consistently and practically applied to existential risk.

### 3.3 Existential Risk: Who is Threatened?

The original definition of the techno-utopian paradigm is not concerned with humans per se. Instead, it is focused on Earth-originating intelligent life, and enhanced posthumans. This is a different inquiry to studying human extinction, and it is not obvious that it should be conducted under the banner of existential risk. The existing species *Homo sapiens* can be approached empirically, offering the opportunity to develop a science of existential risk; by By taking an interest in the future of *Homo sapiens*, scholars can approach existential risk reduction as a communal project, able to engage with the subject of inquiry —existing humans— and consider their individual preferences and visions of what a good future would look like.

## 4. Flawed Definitions, Frameworks and Tools: How Ambiguity and Unverified Assumptions Undermine Research into the Apocalypse

### 4.1 Ambiguous Definitions

This section will look at the problem of defining existential risk. We look at three definitions that can be considered part of the TUA, where an existential risk is one that:

a.     "threatens the premature extinction of Earth-originating intelligent life or the permanent and drastic destruction of its potential for desirable future development"[11];

b.     "threatens the destruction of humanity's long-term potential" (through extinction, unrecovered collapse, or permanent dystopias)[6];

c.     "causes the loss of a large fraction of expected value"[28].

The three definitions share a core feature: they are all fixated on future value. The tragedy to be averted is not the suffering or loss of existing humans, but rather the loss of future value or potential. All the definitions are motivated by future, long-term value. There are also some differences. In his extended typology (see Section 2.1) Bostrom's definition enshrines a particular moral view by specifying desirable futures as technological maturity. The other two definitions are, at least in theory, more abstract and value-agnostic.

By leaving "value" and "potential" undefined, these latter definitions theoretically avoid the charge of existential risk as being a project of a niche philosophical view. According to Ord, our potential should be the entire set of possible futures. He suggests that we should first reduce existential risks to a minimum level to achieve "existential security" before undertaking a "Long Reflection": a patient, collective discussion of what exactly humanity's potential is. There are significant problems with this approach.

First, in practice, value is still expressed in techno-utopian terms. For example, the last chapter of *The Precipice* expands on a vision of humanity's potential: transhumanist space expansion receives ample attention and adoration. Unrecoverable civilisational collapse (a state in which technological progress is not ensured) is described as an existential risk. Here, "civilisational collapse" refers to a permanent reversion back to non-agricultural ways of living. It is not explained why the presence of agriculture, or many of the commonly assumed trappings of "civilisation", such as urbanism, writing, and states (although these rarely came as a coherent package, see Graeber and Wengrow[52]) would increase the likelihood of reaching our potential. For a techno-utopian, it does. For others who value virtue, freedom, or equality, it is unclear why a long-term future without industrialisation is abhorrent: it all depends on one's notion of potential. The definition is seemingly agnostic in the abstract, but in practice there are numerous signals that it expresses the same commitment to total utilitarianism and transhumanism.

Secondly, we need to define what our potential is before we can identify threats to it. How else would we know which risks to address? This is an inherent tension within *The Precipice* since we are supposed to achieve existential security before undertaking the Long Reflection. It is difficult to know if we have achieved existential security if we haven't defined what an existential risk is, since we haven't undertaken the Long Reflection to define our potential. A reasonable counter could be that, in theory, there are certain futures that almost no one would like to live in (such as nuclear winter), and that there may be certain risks (for instance, an asteroid strike) that would take lots of plausibly good options off the table. Extinction may indeed be an outcome which we could assume most people would agree we should avoid. Beyond this point of convergence, there may be far more disagreement on what futures are worth protecting.

Indeed, agreement on the badness of extinction and disagreement over our potential is evident in wider philosophical debates. Philosopher Elizabeth Finneron-Burns argues that extinction is wrong due to the suffering and psychological traumas it could cause, not due to the prevention of millions or billions of people yet to be born[37]. Similarly, an intrinsic end value of humanity can be grounds to want to ensure the long-term survival of humanity, but not the potential for many additional future lives[38]. Others have argued that the badness of extinction is a generation-centred issue and neither the future masses or appeals to the natural lifespan and shape of humanity (there isn't one) provide sufficient grounds[39]. There is a strong case that different value theories should be concerned about extinction. There is not a compelling case that many are compatible with longtermist concerns about the deep future.

Third, the "Long Reflection" and Ord's definition assumes we can reduce risks to humanity's potential without choosing between conflicting values. This is almost certainly not the case. Maybe Ord means that we should reach existential security without closing off any possible futures but retaining option value is presumably restricted to future options that are morally valuable. Indeed, the very notion of existential risk presumes that certain futures, such as dystopias, are to be avoided. We must define what is morally valuable to identify what is dystopian, and we cannot wait until a "Long Reflection" to do so.

Beyond the simpler domain of extinction, things get far murkier. Different plausibly good futures will often involve trade-offs against each other. Steering the world through an age of perils will involve difficult choices. Choices that will frequently have divergent answers depending on one's values. Look no further than the COVID-19 pandemic to see how comparatively smaller crises lead to clashes in values and understandings of what defines a good society. The field should provide clear delineations between risks that were identified as threatening across a broad swathe of value assumptions, and risks that are only threatening given a particular notion of potential. Some scholars may choose to stick with studying extinction risk, rather than trying to specify all possible good futures and the existential threats to them.

Deciding what risks are worth taking and which risks should be taken seriously will need to be a matter of reflection and collective decision-making if it is to respect moral uncertainty and diverse preferences.

Ord writes "If we steer humanity to a place of safety, we will have time to think"[6], but who is "we"? What appears a risk worth accepting to some will not be considered a risk worth taking by others. For instance, slowing technological progress is a risk to a transhumanist who sees our potential in overcoming human biological limitations, but accelerating technological progress may be perceived as a risk by others. In the following sections, we will list several examples of proposed mitigation efforts which show that judgements of strategies to mitigate risk depend on subjective notions of value. Generally, it is easier to identify universal extinction risks than universal existential risks.

Fourth, in practice the attempt at neutrality can end up masking rather than eliminating values from the analysis. Explicit commitments to transhumanist values in Bostrom's definition of existential risk have the advantage of transparency. It is easier to reject or counterbalance a researcher's perspective when their underlying values are clear. Abstract definitions can end up implicitly incorporating moral assumptions. This can happen unbeknownst to the researcher.

Fifth, the definitions are sufficiently ambiguous to render them unfit for use in a rigorous or replicable risk assessment. If we conceptualise an existential risk as the "permanent and drastic destruction" of desirable future development[11,13], human potential[6], or expected value[28], then how severe does the destruction need to be? How significant must be the loss of human potential or expected value? How high does the likelihood of its permanency need to be? Even if a risk is judged sufficiently impactful, how large does the probability of the risk occurring need to be, before it can be considered as a legitimate existential risk? How should we compare such risk being incurred by some action against potential benefits of said action? Should risks based on speculation or thought experiments or naive technological extrapolation be treated seriously (see Section 5.2)? These decision-relevant ambiguities have not yet been clarified.

These are not minor, theoretical quibbles. Empirical and moral assumptions determine what is considered an existential risk and what mitigation efforts are recommended. Whether or not these choices are considered reasonable doesn't depend on a replicable framework. Instead, it relies on whether the judge shares the same assumptions.

An example: Ord[6] considers Artificial Intelligence (AI) to be the biggest contributor to existential risk within the next 100 years. The author's probability estimate relies on a survey of machine learning researchers[53], a study with questionable methodological value for determining whether AI is an existential risk[54,55]. The policy recommendations for mitigating such risks in *The Precipice* support R&D into aligned artificial general intelligence (AGI), instead of delaying, stopping or democratically controlling AI research and deployment.[viii]

---

[viii] Note this unwillingness to stop or delay progress in technology or research is interestingly often not observed in biorisk research in ERS. It is unclear why this is the case. One explanation could be that banning or stopping technologies and practices in biorisk is seen to be more feasible than doing the same for other areas. We have yet to see a convincing argument as to why this should be the case. Another explanation could be that this is about desire, not tractability. AI is central to the techno-utopia vision of the future. Gain-of-function experimentation is not. This suggests that the techno-determinism of the TUA is prescriptive, not descriptive (see section 4.3 for further discussion).

These policy recommendations were echoed in the "Future Proof" report by the Centre for Long-Term Resilience[46], which was aimed at UK policymakers. This recommendation seems to assume a kind of technological determinism (see below), or an implicit advocacy for building advanced technology for instrumental purposes. Either explanation echoes the TUA and leads to recommendations that support a particularly existentially risky course of action: developing advanced AI. Despite an explicit acknowledgment that AI could be a major contributor to risk and that slowing, delaying or halting development can help avoid the risk, the book recommends R&D, rather than e.g., citizen surveys, moratoria, or transparency measures.

What appears to be a risk worth accepting to some will appear to be a risk not worth taking to others. Take for example Bostrom's "Vulnerable World Hypothesis"[17], which argues for the need for extreme, ubiquitous surveillance and policing systems to mitigate existential threats, and which would run the risk of being co-opted by an authoritarian state.[ix] It is a solution that some may find appealing and others appalling. Without a deliberation between different moral views, it should not be assumed that this risk is acceptable. Given disagreements about risk-taking, the field needs to ask who gets to put forward recommendations and who gets to choose between them (see Section 6).

Many definitions face the challenge that they are too abstract to allow for robust, replicable analysis, but the TUA's definitions of existential risk are particularly faulty. These definitions conflate the study of global catastrophe or human extinction with that of the longtermist ethics of existential risk. The question of what futures are worth taking which risks for will always rely heavily on a notion of value. For this reason, we suggest scholars consider separating the following areas:

- *Extinction Ethics*: the study of the ethics of human extinction; the badness or goodness of human extinction given different ethical considerations.

- *Existential[x] Ethics*: the study of the ethical implications of different societal forms. This includes arguing (under different ethical assumptions) for the relative goodness of different futures and how they relate to human potential – both in the deep future and today. This in turn provides ground for defining what would constitute an *existential* risk.[xi]

---

[ix] While some may choose to read Bostrom's discussion as a merely speculative proposal for a hypothetical scenario, it exemplifies a willingness to consider, and even justify, measures that would normally be deemed unconscionable in the name of averting supposed existential catastrophes. Moreover, the fact that the essay was published in a journal for policy (with policy implications considered), rather than philosophy, suggests that the discussion should not be considered as entirely theoretical and not actionable. Additionally, intentional outreach through a TED Talk (with over 2.5 million views), the Sam Harris podcast, and a piece in Aeon all stress this aspect of the article. These actions signal that the argument is not intended to be an innocuous thought experiment.

[x] The term *existential ethics* is already sometimes used in relation to the philosophy of existentialism. We are aware of this; however, it seems difficult to move away from the use of the existential given the field's reliance on the term existential.

[xi] An easier initial step here may be to specify dystopias that most value theories would say we should avoid, rather

- *Catastrophic Risks*: the study of contributors to the occurrence and probability of global catastrophic events.

- *Extinction Risks*: the study of contributors (which include global catastrophic events) to the probability of human extinction.

While each inquiry can have both empirical and theoretical elements, the latter two lend themselves to scientific risk analyses, while the former two are more philosophical inquiries by nature. All the existing definitions currently conflate all of these. For many analyses, this is unnecessary and counterproductive. The media repeatedly makes the mistake of equating Ord's estimated 1 in 6 chance of existential catastrophe with extinction risk[56,xii] but these are meant to be drastically different concepts. The study of extinction and existential (or longtermist) ethics does not need to be resolved to scientifically study risks, and it requires a different set of skills and procedures (see Section 6). There will be some necessary and fruitful areas of overlap. For instance, ensuring that measures to prevent and mitigate risks are proportional and not dangerous, we will need to have some debate on existential ethics.

Risk analysis will always be at least partly subjective and value-laden. Yet, it can be made more objective and scientific through precise definitions, the transparent statement of assumptions, and where possible, separating risk assessment from the study of extinction and exisential ethics.

### 4.2 Arbitrary Categorisations

ERS currently lacks a framework or methodology to categorise risks consistently, comprehensively, and rigorously. The TUA does not provide such a framework. It purports to distinguish between existential and catastrophic risks, but the distinction is hazy and arbitrary.

ERS currently distinguishes between existential risks and "global catastrophic risks". All existential risks are global catastrophic risks, but not all global catastrophic risks are existential risks. Definitions of global catastrophic risks proposed in the literature vary but tend to focus on a significant global loss of human life (such as a loss of 10%)[20]. Most include the great catastrophes of the past, such as the Black Death.

Under the earliest TUA definition, a catastrophe that does not jeopardise the attainment of technological maturity is assigned comparatively little moral consideration. This is due to the belief that such disasters have not influenced our long-term fate, and thus do not constitute existential risks,[xiii] and that limited resources would be wasted if they were directed towards global catastrophes that did not threaten technological maturity[11]. This extreme prioritisation of existential risks downgrades the importance of addressing other global catastrophes. What under Bostrom's view would plausibly be considered "feel-good projects of suboptimal

---

than develop a widely shared notion of potential.

[xii] We owe this point to Joshua Teperowski Monrad.

[xiii] They are, in Bostrom's words, "mere ripples on the surface of the great sea of life" (Bostrom, 2002).

efficacy", and what falls short of "efficient philanthropy"[11], could in fact very much be worth the attention of those who study human extinction.

Ord[6] presents a more moderate version of this original TUA framework. He considers it justified to spend some resources on large catastrophes, because those catastrophes could indirectly amplify (but not directly cause) existential risk. His framework still draws clear distinctions between existential and catastrophic risks framing some as direct "risks" and others as indirect "risk factors", prioritising the reduction of direct risks.

We lack the necessary understanding of how human society and the Earth system operate to make such neat, surgical distinctions. Whether a single global catastrophe can or would fundamentally alter the trajectory of humanity is one of the great unanswered questions of history. For now, we struggle with forecasting GDP even a year ahead[57] and in fact reliably make incorrect forecasts about population growth, despite knowing all essential variables[58]. In a complex adaptive system it may be impossible to forecast how small changes will affect the longue durée. Neat distinctions between GCRs and existential risks presume a level of systemic knowledge and certainty we currently do not possess. The TUA thus far does not offer the tools to make any fine-grained, credible separation between existential and non-existential global catastrophes. It has not offered explanations of how such events affect extinction and existential risk.

This is not to say that some persistent, long-term, societal trends cannot be identified and understood. For instance, the historian Walter Scheidel has put forward a compelling empirical case that (intra-country) wealth inequality increases inexorably until a great leveller (a state collapse, pandemic, revolution, or mass mobilisation warfare) resets the playing field[59]. Investigating these trends will be critical to foreseeing how catastrophic risks are produced and could unfold deep into the future. Such an analysis doesn't require or justify the crude split between global catastrophes and extinction risks.

Under the TUA, an existential risk is understood as one with the potential to cause human extinction directly or lead us to fail to reach our future potential, expected value, or technological maturity. This means that what is classified as a prioritised "risk" depends on a threat model that involves considerable speculation about the mechanisms which can result in the death of all humans, their respective likelihoods, and a speculative and morally loaded assessment of what might constitute our inability to reach our potential.

Imagining pathways to human extinction (kill mechanisms) invariably requires some creativity and speculation. This has meant that some areas of risk (e.g. AI) which are not as empirically constrained are often prioritised above others for which we have far more empirical data (e.g. climate change). This has led some (non-peer-reviewed) publications (including a Google Doc) to conclude that climate change is not an existential risk[60,61,xiv].

This is the wrong question to ask. Adding up the predicted impacts of a selection of hazards and asking, "will the total of these kill everyone?" is a simplistic and ineffective way of

---

[xiv] Halstead does not provide a definition of existential risk, so it is unclear what this means.

conducting risk analysis. This is not how risk unfolds in reality: hazards interact with networks of societal vulnerabilities and responses as well as each other, and can trigger cascading failures[62]. We need to consider different pathways and ways in which climate change (or any other source of risk) can contribute to the overall level of extinction or catastrophic risk we face[62,63]. The question of "is this an existential risk?" is naive. We should instead ask: in a given world-state (with structure, vulnerabilities, and the capacity for change) how much will a given process or event increase the overall likelihood of human extinction, and what are the plausible[xv] pathways for it to contribute to extinction risk?

A field looking for the *one hazard to kill them all* will end up writing science fiction. More speculative risks are prioritised because a seemingly more complete story can be told and speculative mechanisms by which AI could kill every human can seemingly not yet be ruled out.

In practice this could be addressed by lowering the threshold of what risks should be treated as relevant to extinction and include more of what Ord calls "risk factors", such as those commonly thought of as global catastrophic risks (GCRs). Global catastrophes and responses to GCRs can give vital insight into vulnerabilities that should be mended and resilience factors that should be enhanced.

The distinction between "direct" and "indirect" risk factors also depends on speculation. Direct risks appear to be those for which we can tell a story about how they might "directly" (presumably limited to third or fourth order effects) lead to the extinction of *Homo sapiens*. Unaligned AI, for example, is often considered a direct risk, but the story about extinction from AI is far from complete[xvi]. Strong expert disagreement regarding risks from AI[53,54] is testament to how debatable the empirical foundations of "direct" risk pathways of AI still are. Additionally, risks originating from AI could come in many forms, each relying on speculation about different kill mechanisms and assumptions about the nature and use of a system that has never been built. Surely, not all these pathways are equally direct, and yet, AI is prioritised as a seemingly homogenous hazard-cluster across key texts within ERS[3,6].

A risk perception that depends so strongly on speculation and yet-to-be-verified assumptions will inevitably (to varying degrees) be an expression of researchers' personal preferences, biases, and imagination. If collective resources (such as research funding and public attention) are to be allocated to the highest priority risk, then ERS should attempt to find a more evidence-based, replicable prioritisation procedure.

### 4.3. Simplistic Risk Models

---

[xv] Defined as being consistent with our background knowledge (Betz, 2016)
[xvi] "The case for existential risk from AI is clearly speculative. Indeed, it is the most speculative case for a major risk in this book." (p. 149, Ord, 2020).

### 4.3.1 Complex vs. Crude Risk Assessments

Risk assessment has evolved dramatically in past decades. Scholars now commonly analyse systemic risk (the ability for a single disruption to cascade into systems failures)[64], how risks can cascade across borders and sectors[65], and how failures in critical systems can synchronise and reinforce each other[66]. This has led to new forms of complex risk assessment, particularly in climate science and disaster risk reduction. The Intergovernmental Panel on Climate Change (IPCC) sees risk as composed of vulnerabilities, hazards, and exposures, as well as response risks[67]. Similarly, others have suggested that a complex risk assessment needs to consider four determinants of risk (hazard, vulnerability, exposure, response) as well as how risks link and cascade. Understanding the common drivers across each of these determinants is critical to mitigation efforts[68,xvii].

How we assess risk is fundamental to what we consider as an existential or catastrophic risk. For instance, Ord focuses on super volcanoes as a potential existential catastrophe. However, lower magnitude volcanic eruptions could have catastrophic impacts due to their cascading effects and the vulnerable nature of critical infrastructure systems[69]. Similarly, stratospheric aerosol injection[xviii] does not appear to pose direct risks that would classify it as a global catastrophic threat. Yet, this depends on how it is deployed, the world in which it operates, and the level of warming it is masking. If another calamity, such as a volcanic eruption, solar flare, or nuclear war, destroys the mitigation system for a prolonged period, the ensuing 'termination shock' (rapid global warming over a short timeframe) would likely result in catastrophic effects[63]. A more complex risk assessment will be more difficult to do, but it will be more accurate, realistic, and informative.

Most existential risk texts take a simpler, hazard-centric approach. They tend to focus on a few selected hazards: biologically engineered pandemics, Artificial General Intelligence (AGI), nuclear war, climate change, and asteroid strikes[2,6]. As currently framed, TUA equates risk with hazard and ignores the wider literature on risk assessment in fields such as disaster risk reduction. It is also unclear how the TUA suggests systematically clustering, prioritising, or analysing these hazards: current attempts rely on simplistic categories of "Natural", "Anthropogenic", and "Future" hazards[6] or presenting the selected hazards as the ones worth discussing without explanation. There have been some recent attempts to provide alternative frameworks[18,19,70] but these have found little application thus far, and still do not consider response risks and many other relevant areas. Research efforts are often split across the lines of these different hazards. Working across them as part of a more complex risk assessment could offer novel insights.

---

[xvii] In disaster risk management even more risk components have been proposed, such as capacity. We highlight this formulation due to the prominence and rigour of the IPCC.
[xviii] The injection of aerosols into the stratosphere to reflect sunlight and mitigate some of the effects. of climate change.

### 4.3.2 Technological Determinism

The choice to structure risk assessment this way has not been explained or defended. It may have been chosen due to an implicit techno-determinist threat-model: the TUA often appears to assume an exogenous threat model in which existential hazards naturally and apolitically arise from inevitable and near-autonomous technological progress. The TUA rarely examines the drivers of risk generation. Instead, key texts contend that regulating or stopping technological progress is either deeply difficult, undesirable, or outright impossible[2,6,11]. Bostrom proposed a "Technological Completion Conjecture": if technological developments do not cease, then all important, basic technological capabilities will be obtained in the long run[71]. Others offer a more sophisticated view, in which military-economic competition exerts a powerful selection pressure on technological development. This "military-economic adaptionism" constrains sociotechnical change to deterministic paths. Technologies that gift a strong strategic advantage will almost certainly be built[72]. Many in the related Effective Altruism community disregard controlling technology on the grounds of a perceived lack of tractability[73].

Whether it is technological determinism, the more nuanced military-economic adaptationism model, or concerns around tractability, the result is the same: regressing, relinquishing, or stopping the development of many technologies is often disregarded as a feasible option.

The proposed alternative is "differential technological development": speeding up and slowing down different technologies to ensure they occur in the safest order possible. Why this is more tractable or effective than bans, moratoriums and other measures has not been fully explained and defended (see Section 5.1.1 for further discussion). This could be interpreted as hard-nosed pragmatism, an argument that we can stop technologies from being built, but it will not be an efficient or prudent use of resources. Again, a compelling analysis has not been made for why this is the case, and in practice this ends up looking identical to technological determinism. The irony is that if the world is locked into the development of dangerous technologies, then we are already in a "lock-in scenario" so dreaded by many within the TUA[6]. In the eyes of the TUA, the range of future options available to humanity is already greatly restricted.

It is unclear whether technological determinism in the TUA is descriptive or prescriptive. It could be a genuine belief that controlling technology is infeasible. It could also be that under the TUA unabated technological progress is vital to achieve technological maturity and avoid existential risk.

In any case, the assumption of technological determinism leads scholars to focus on hazards, rather than, say, exposure, maybe because there appears to be no point in trying to reduce humanity's exposure to a technology since the development of the technology is assumed inevitable. It is then merely a question of whether benevolent technologies are built first. Attempts to change political or economic drivers of different risks get less attention. This unstated threat model leads to sharp divergence from modern developments in risk analysis towards a crude hazard-centrism.

Importantly, assumptions around technological determinism are highly contested. Indeed, technological determinism is largely (for better or worse) derided and dismissed by scholars of science and technology studies[72]. We have historical evidence for collective action and coordination on technological progress and regress. One example is weather modification. Early attempts were made by the US during the Vietnam War to use weather modification technologies to extend the monsoon season and disrupt enemy supply chain[74]. The introduction of the 1976 Convention on the Prohibition of Military or Any Other Hostile Use of Environmental Modification Techniques (ENMOD Convention) seems to have successfully curtailed further research into the area[75]. The assumptions on technological progress must be thoroughly examined, empirically and theoretically, before they should be used to determine policy and mitigation actions.

There are enough exceptions to doubt that any strategically powerful technology will be due to competition between (largely) rational actors (usually states). Many important technologies such as glass and steam engines were used for ceremonial purposes for centuries before being redirected towards practical purposes[76]. During the early industrial revolution water mills were more reliable and efficient than coal-fired steam engines. The latter were adopted not because of their inherent superiority, but because they could be located in urban areas with a large and desperate population, which appealed to early capitalists[77].

Any approach to existential risk will struggle to find frameworks that are comprehensive yet elegant and practical. It will need to be transparent about its empirical assumptions, including on how risks are created. For now, the hazard-centrism and opaque technological determinism of the TUA provides a framework that is overly simplistic, unduly curtails the available mitigation options, and provides no compelling method to understand or address the common drivers behind risk determinants. It is inadequate for the grand challenge of understanding and mitigating extinction risks.

There is room for different empirical worldviews, different frameworks, and different moral positions to be considered in the study of existential risks. We do not at all recommend that hazards should no longer be studied. Similarly, speculation will always be a part of assessing unseen risks, but a science of extinction should adopt frameworks which minimise the need for this aspect. Different frameworks will make different predictions about what policy and research efforts appear to plausibly reduce risk.

### 4.4 Inappropriate Translations of Theory into Practice

Mitigating existential risk requires decision-making under uncertainty. Decision theory in the context of ERS and longtermism is an active area of research. For now, we want to caution against applying idealised decision-theoretic results to the evaluation of risky choices in practice. This is because empirical uncertainty can affect the applicability of results that hold in theory and because expressing subjective notions of risks and benefits numerically can provide a false sense of certainty.

Take for expected value (EV), defined as the value of the outcome multiplied by the probability of it occurring. The TUA extensively uses expected value calculations to justify its own approach and prioritisations. For example, Bostrom argues that existential risk mitigation should be prioritised over other altruistic acts: if there is just a 1% chance of $10^{54}$ people coming to exist in the future, then "the expected value of reducing existential risk by a mere one billionth of one billionth of one percentage point is worth a hundred billion times as much as a billion human lives"[13]. According to Bostrom, "even the tiniest reduction of existential risk has an expected value greater than that of the definitive provision of any "ordinary" good, such as the direct benefit of saving 1 billion lives"[11]. Elsewhere, Millett and Snyder-Beattie use EV to argue for reducing risks from biological pathogens[78].

While EV is a useful theoretical tool in a range of contexts, in practice, it is hard to apply rigorously when working with the generally low and highly uncertain probabilities characteristic of existential risks. The TUA applies expected value theory to the very areas where it faces the most pitfalls, that is, situations of deep uncertainty and low information about probabilities[79]. Ord attempts to estimate probabilities of existential catastrophes caused by various hazards for communicative purposes, most notably the aforementioned one in six figure[6]. While this has been successful in terms of public communications (the estimate has been widely reported in the press), it is unclear how to evaluate the accuracy of these estimates or whether his methodology for arriving at them is sufficient to warrant the sense of scientific credibility that numbers inevitably imply to the lay person. In addition, to evaluate the EV of mitigating an event, one must decide upon a particular conception of value. Thus, personal intuitions and non-representative moral preferences are at risk of being captured and made to appear objective by numbers. How can policy recommendations be considered robustly good and replicable if they are evaluated on subjective assessments of probabilities and hidden ethical assumptions? EV is still unsuitable to be relied upon in practice for estimating human impact on the value of the long-term future[80,81].

Furthermore, EV and decision theories more widely are affected by Pascal's Mugging[82,83] as well as what has been called *fanaticism*[84]. We know of no pragmatic and consistent response to those challenges yet. Fanaticism describes how we may be required to put considerable effort into mitigating terrible events with an arbitrarily miniscule probability of occurring. Similarly, Pascal's Mugging describes how vast or near-infinite quantities of value can overwhelm even the most minuscule probabilities: the term comes from a thought-experiment in which Pascal is conned by a self-proclaimed wizard who promises to magically grant 1,000 quadrillion happy days in exchange for his wallet[83]. The probability that the grifter is a powerful wizard is infinitesimally low, but not zero, and outweighed by the expected utility of so many happy days. The Pascal's Mugging problem arises when applying EV to existential risks as defined within the TUA. Any risk that could prevent technological maturity, no matter how small, should be taken seriously and acted on due to the sheer amount of expected value at stake. Scholars have suggested that Pascal's mugger returns to swindle another unsuspecting victim, but this time using existential risk studies and longtermism[85].

Both these challenges are usually evaded by claiming that hazards like AGI have an unambiguously high enough probability of occurring this century to merit considerable

action[xix]. This only side-steps the question in the case of AGI in particular (assuming that these doubtful estimates can be trusted) and does not tell us whether highly unlikely or speculative risks should be acted on. There are good reasons to defend fanaticism[84] and further theoretical work might resolve the challenges presented here, but while the integration of theory and empirical work is still ongoing, we should consider drawing pragmatic lines between mere speculation and risks humanity should significantly focus on. For instance, a simple threshold or plausibility assessment[86] could protect the field's resources and attention from being directed towards highly improbable or fictional events.

### 5. The Risks of Studying Existential Risk

How could the study of global risks and longtermism contribute to catastrophe? The worse-case outcome is not that existential risk remains unaffected, or that resources are wasted on incorrect speculations (although these are problems). Instead, it is that these risks are aggravated by research into them. This issue must be addressed for any approach to ERS. Unfortunately, the TUA appears to be particularly prone to both ignoring response risks and aggravating them.

### 5.1 A Risky Road to Safety

Mitigating risks incurs a risk of its own. As noted earlier, this is a fundamental part of sophisticated risk assessments, including those used by the IPCC[67,68]. Not all approaches incur the same risks. While we will not compare the techno-utopian approach against alternatives in this paper, we think the TUA is prone to an especially high level of response risk.

The zealous pursuit of technological development, according to proponents of the TUA, accounts for the vast majority of risk over the coming centuries. The risk of human extinction from natural hazards is likely low, with an upper-bound of less than one in 14,000[87] and a best guess of around 0-0.05% per century[6,xx]. In contrast, several scholars of existential risk place the likelihood of an existential catastrophe far higher at 1/6[6], or >1/4[13] over the coming century. Rees puts the chance of collapse or extinction by 2100 at 1/2[2]. This discrepancy is mainly due to anthropogenic risks arising from climate change, nuclear weapons, synthetic biology, and artificial intelligence (AI).

If the lion's share of extinction risk stems from emerging technologies, why do we rarely ask how to stop dangerous developments? This option is usually considered infeasible[6], or outright impossible[17]. This may be due to the exogenous threat model and technological determinism of the TUA. Since halting the technological juggernaut is considered impossible, an approach of differential technological development is advocated[6,13]. This involves trying to

---

[xix] Based on personal communications with several scholars in the field.
[xx] These figures are not water tight: they assume that the vulnerability of humanity in the Palaeolithic is identical to today.

develop beneficial and protective technologies (aligned artificial intelligence[xxi] is often used as an example) first, before proceeding to riskier options.

It is not clear how scholars plan to reliably determine which non-existing technologies will be more or less risky years or decades in advance. Even if we did have such a refined vision of the future, it is unclear why a precise slowing and speeding up of different technologies (which are interlinked and presumably require a set of fine-tuned regulatory tools) across the world is more feasible or effective than the simpler approach of outright bans and moratoriums.

More importantly, in the TUA the stark choice between one of only two destinies — technological maturity or existential catastrophe — is a fait accompli. The path to techno-utopia appears to be the only one available, despite its risks. From a techno-utopian perspective, a failure to build these dangerous, powerful technologies is an existential risk. Bostrom, aware of the tension arising from recommending the (albeit careful) development of technologies, warns: "We should not blame civilization or technology for imposing big existential risks. Because of the way we have defined existential risks, a failure to develop technological civilization would imply that we had fallen victims of an existential disaster. […] Without technology, our chances of avoiding existential risk would therefore be nil"[13]. The TUA dramatically restricts the options available for avoiding existential catastrophe.

Furthermore, pursuing a techno-utopian future is dangerous and may come with considerable cost. It has already been noted that the attempt at colonising space and an expansion of technological capabilities could end in catastrophe if it foments a new arms race and large-scale warfare[88]. Upgrading the human body could construct a biological caste system, where, an enhanced, genetic elite could oversee a subjugated, unenhanced, "inferior" class[89,90]. These are not far-flung speculations. Researchers already actively monitor, evaluate and debate near-term bio-engineering enhancements and their ethical implications[91,92]. Enhanced inequality would not only be unjust, but also amplify many social ills[93,94]. Similarly, a horizon scan under the WHO Science and Research Division has noted that the pursuit of advances in the life-sciences could produce many technologies that could be easily misused to cause harm. These range from using bioregulators for the delivery of bioweapons to the use of deep learning algorithms to identify novel biological pathogens[95].

Existential risk does not need to be defined in reference to technological maturity, nor does it need to be accompanied by these response risks of accepting or even speeding up disruptive technologies. A different vision for a good future could lead to dramatically different policy recommendations. A less determinist view of technological change and pessimistic view of political change would open up a plethora of other interventions. A democratic approach to ERS will provide ample room for different moral and empirical assumptions to affect the assessment, discussion, and negotiation of collective risk-taking. Other paths and approaches may be more risk-averse and must be explored if humanity wants to safely reduce existential risk.

---

[xxi] That is, a superintelligent AI system that is aligned with human values and will aid rather than harm humanity.

## 5.2 High Stakes: Existential Exceptions and a Risk-Averse Approach to Existential Risk

### 5.2.1 The Stomp Reflex

There is a long history of security threats being used to enable draconian emergency powers. Emergency powers are intended to be conservative: to protect existing legal and political structures in a period of tumult. The logic is that drastic times call for drastic measures. To protect institutions, emergency powers allow governments to disregard existing laws and exempt themselves from judicial or democratic restrictions and oversight. Rather than protect, such measures are often abused to erode and transform fundamental political structures; when trying to centralise and extend state powers, fear is a powerful justification. The larger the fear, the easier it is to justify more potent emergency powers. If the perceived threat is human extinction, then the measures could be extreme.

Recent examples abound. The Patriot Act, adopted by the US just 45 days after the 9/11 attack, allowed for a range of draconian actions, including indefinite detainment of migrants without criminal prosecution. The provisions broadly underpin the current US surveillance network, most notably through the NSA's PRISM program, which Edward Snowden exposed in 2013. The *War on Terror* became a useful cover for the creeping power of the US security apparatus. This despotic drift is not a purely historical threat. Clauses for states of emergency have spread over past decades[96,97] and 2020 marked the highest ever use of such measures.

The prolific use of emergency powers can lead to the creation of a *state of exception* in which the sovereign transcends the regular rule of law. Temporary measures become permanent, and spill into the operation of the legal system[98]. The transition from the Roman Republic to the Roman Empire, the fall of the Weimar Republic in the 1920s and 1930s into the Nazi regime, and many other political declines were underpinned by the normalisation of emergency powers[99].

The irony is that emergency powers rely on an inaccurate understanding of human nature and disaster risk. Emergency powers inevitably empower those atop hierarchies, despite abundant evidence from disaster risk reduction and other fields that while mass panic is a myth, the risk of elite panic and elite co-option of catastrophes is real[100–103]. There is even evidence that such a response worsens crisis. One study of natural disasters found that the larger the number of emergency provisions used by an executive, the higher the fatalities (controlling for disaster severity and size)[104]. This is a "*Stomp Reflex*": governments using emergency powers to reassert and veil systems of authority. Such a response is counter-productive and ultimately shifts power into the shadows, away from transparency and public accountability[105].

Existential risk is the perfect excuse for enacting the Stomp Reflex. Indeed, catastrophic hazards such as nuclear weapons have already been used to justify anti-democratic shifts. In the US, the accumulated nuclear stockpile and threat of sudden war justified a profoundly autocratic move: a single individual the President — was given the ability to launch nuclear attacks. Richard Nixon once boasted "I can go into my office and pick up the telephone and in twenty-five minutes seventy million people will be dead". As sociologist Elaine Scarry has argued, the nuclear decision-making apparatus violates constitutional rights, the deliberative nature of democracy, and any social contract. The world lives in the shadow of a "thermonuclear monarchy" rather than a democracy[106].

Nuclear weapons also saw a revolution in secrecy in the US. The threat of thermonuclear war was used as a justification to construct unprecedented levels of secrecy in the military and intelligence communities. These were of dubious efficacy in preventing the spread of nuclear weapons, but they did have the effect of eroding transparency and democratic control over the military industrial complex[107].

The best empirical example of policy responses to an existential threat do not inspire confidence. In the US at least, the threat of thermonuclear war spurred dramatic reforms that made for a less democratic and open state, but not necessarily a safer one.

### 5.2.2 Survival Through Security, Surveillance and Suppression

Any approach to understanding and mitigating existential risks runs the risk of becoming *securitised*. Securitisation refers to a discursive manoeuvre that moves an issue from the arena of normal politics to that of national security, making it more likely to permit emergency powers and be placed under the control of unelected military and intelligence officials. Moves towards thermonuclear monarchy and elevated secrecy were largely underpinned by neorealist foreign policy and game theory developed in such a context[xxii]. This is not to say that all securitisation approaches are equally dangerous; some are far more likely to enable authoritarian responses.

There are reasons to expect that the TUA is particularly vulnerable to misuse. As philosophers such as Peter Singer and Phil Torres have noted, if the world is viewed from the TUA's lens of existential risk, then we run the risk that almost any action is justified if it is believed to improve our chance of surviving to expand beyond Earth[108,109]. Problems which are not considered to be an existential risk dwindle into irrelevance, as other values are sacrificed on the altar of expected astronomical value.

This is not to say that most believers of the TUA are intent on using it to justify morally abhorrent actions, nor that they are unaware of these weaknesses. Rather, we argue that the basic logic of protecting a high-tech future of astronomical value could be easily co-opted. Marx never intended for communism to justify brutal dictatorships. Nonetheless, it was easily twisted by Stalin and others to do so.

Scholars of existential risk have already shown some proclivity for invoking security, whether it be for "existential security"[6] or "epistemic security"[110]. Extreme emergency responses have also been raised. Bostrom's "Vulnerable World Hypothesis"[17] identified the combination of ubiquitous surveillance, preventative policing, and global governance (understood to be "a

---

[xxii] It is worth noting that both neorealist assumptions about international relations and game theoretic approaches appear frequently in research under the TUA.

world government; a sufficiently powerful hegemon or a highly robust system of inter-state cooperation")[111] as a comprehensive agenda to protect against possible technological hazards that could devastate civilisation. He proposes a typology of four potential threats: "easy nukes" (readily accessible and easy to use weapons of mass destruction), "safe first strike" (the ability to safely destroy others with impunity), "surprising strangelets" (experiments that could harbour an unforeseen, or foreseen but low-probability catastrophe); and those in which the accumulation of minor damages by individuals eventually accumulate into global catastrophe.

Bostrom's preferred solution — extreme preventative policing and widespread surveillance — could involve the mandatory use of ironically-named "freedom tags" fitted with multiple cameras and microphones to continuously track individual behaviour. These would be distributed to all citizens and monitored by state employees — "freedom officers" who themselves are watched by artificial intelligence to prevent misuse — who can order preventative interventions using drones or police[17].

Both the journal paper (published in the journal *Global Policy*), as well as public-facing spin-off articles[111] about the Vulnerable World Hypothesis feature clear policy recommendations. A box in the 2019 paper titled "Policy Implications" includes the recommendations that dealing with "black balls" (technological innovations that by default destroy the world) would require "a system of ubiquitous real-time worldwide surveillance. In some scenarios, such a system would need to be in place before the technology is invented"[17]. The public facing article from 2021 asks: "If you find yourself in a position to influence the macroparameters of preventive policing or global governance, you should consider that fundamental changes in those domains might be the only way to stabilise our civilisation against emerging technological vulnerabilities". It is not difficult to foresee how such ideas could provide grounds for aspiring autocrats to subvert democratic institutions in the face of global threats.

Bostrom[13], also includes a discussion of pre-emptive strikes. He describes the responsibility and need for nations to (on some occasions) enact pre-emptive, unilateral infringements of sovereignty. If extinction threatening technologies (he imagines biosphere-destroying nanobots) are not controlled under international treaties, "the mere decision to go forward with development of the hazardous technology […] must be interpreted as an act of aggression" and would justify pre-emptive infringement of national sovereignty[13].

There is a clear danger in authoritative recommendations based on speculative thought experiments. Scholars using the TUA providing recommendations for surveillance and pre-emptive measures in the name of avoiding catastrophe could contribute to birthing the very dystopias they fear.

There is little evidence that the push for more intrusive and draconian policies to stop existential risk is either necessary or effective. It is empirically dubious to think that we cannot halt or delay the development or spread of dangerous technologies. Nor is it convincing that surveillance measures would prove effective. We have little to no evidence that the use of mass surveillance has been effective at preventing terrorist attacks in the US[112,113]. Moreover, the main creators of such hazards —the *Agents of Doom* — are often the very people who control the surveillance apparatus: military industrial complexes, enormous technology firms, and powerful states[114]. At worst, the knee-jerk reaction of surveillance and preventative policing to prevent a speculative calamity could simply create one of its own: entrenched authoritarianism.

The obvious option to discontinue certain technological developments — if we assume that the Vulnerable World Hypothesis is true — is considered "hardly realistic" and "extremely costly, to the point of constituting a catastrophe in its own right"[111]. Ord too warns of so-called "desired dystopias", in which an ideology (or manipulation and surveillance) has corrupted our choices to the extent that, for example, we "completely renounce further technological progress"[6].

Under the TUA we appear to be trapped. We either develop technologies which carry immense risk, such that ubiquitous surveillance becomes necessary, or we cease technological development only to manifest the existential risk of failing to reach a technologically mature utopia. This trap is an idiosyncratic feature of the TUA.

There are more options for humanity than merely picking between two highly risky paths. Ord confidently asserts that ceasing any further technological development would "ensure our destruction at the hands of natural risks"[6], but we have seen no convincing analysis that shows we could not safeguard our survival with current technologies and re-directed resources. Indeed, it is unclear why we cannot develop technology to address threats from asteroids and supervolcanoes without indulging in the entire range of dangerous inventions. The opinion that all technological progress must continue at all costs is a dangerous one.

Scholars of existential risk need to be vigilant to response risks, and risk-averse in their own suggested interventions. This means we must be truthful about uncertainty, consider the worst-case outcomes of our actions, and verify the acceptability of proposed interventions by subjecting them to democratic oversight.

### 6. Democratising Risk

There is an intimate and neglected relationship between existential risk and democracy. Democracy must be central to efforts to prevent and mitigate catastrophic risks. It is also an antidote to many of the problems manifest in the TUA. Do those who study the future of humanity have good grounds to ignore the visions, desires, and values of the very people whose future they are trying to protect? Choosing which risks to take must be a democratic endeavour.

We understand democracy here in accordance with Landemore as the rule of the cognitively diverse many who are entitled to equal decision-making power and partake in a democratic procedure that includes both a deliberative element and one of preference aggregation (such as majority voting)[45,115]. Decision-making procedures are not either democratic or non-democratic, but instead lie on a spectrum. They can be more or less democratic, inclusive, and diverse.

We posit three reasons for why we should democratise research and decision-making in existential risk: the nature of collective decision-making about human futures, the superiority of democratic reason, and democratic fail-safe mechanisms.

Avoiding human extinction, or crafting a desirable long-term future, is a communal project. Scholars of existential risk who take an interest in the future of *Homo sapiens* are choosing to consider the species in its entirety. If certain views are excluded, the arguments for doing so must be compelling.

Democracy will improve our judgments in both the governance and the study of existential risks. Asking how our actions today influence the long-term future is one of the most difficult intellectual tasks to unravel, and if there is a right path, democratic procedures will have the best shot at finding it. Hong and Page[116,117] demonstrate both theoretically and computationally that a diverse group of problem-solving agents will show greater accuracy than a less diverse group, even if the individual members of the diverse group were each less accurate. Accuracy gains from diversity trump gains from improving individual accuracy. Landemore[115], builds on this work to advance a probabilistic argument that inclusive democracies will, in expectation, make epistemically superior choices to oligarchies or even the wise few. This is supported by promising results in inclusive, deliberative democratic experiments from around the world[118]. In the long run, democracies should commit fewer mistakes than alternative decision-making procedures. If this is true, it should improve the accuracy of research efforts and decision-making. We are more likely to make accurate predictions about the mechanisms of extinction, probable futures, and risk prevention if the field invites cognitive diversity, builds flat institutional structures, and avoids conflicts of interest.

There are many ways to consider the interests of the many. Democratic assemblies could allow global citizens to deliberate about the futures they prefer, citizens could be surveyed, and the field of ERS itself could be diversified. At the moment, the field is, as many academic disciplines are, unrepresentative of humanity at large and variably homogenous in respect to income, class, ideology, age, ethnicity, gender, nationality, religion, and professional background. The latter issue is particularly true of existential risk, which, despite being an inherently interdisciplinary endeavour, is at the highest levels dominated by analytic moral philosophers. We need to be vigilant to what perspectives are not represented in the study of existential risk. An awareness of bias will go some way towards mitigating its negative effects. To get close to replicating the cognitive diversity found among humans, we must begin by inviting different thinkers with different values and beliefs into the field.

Democracies can limit harms. Any approach to mitigating existential threats could create response risks, and the TUA seems particularly vulnerable to this. Despite good intentions and curiosity-driven research, it could justify violence, dangerous technological developments, or drastically constrain freedom in favour of (perceived) security. If we hope to explore ideas but minimise harms, democracies can be used to moderate the measures taken in response to harmful ideas. It seems, for example, vanishingly unlikely that a diverse group of thinkers or even ordinary citizens would entertain the idea of sacrificing 1 billion living, breathing beings for an infinitesimal improvement in reaching an intergalactic techno-utopia. In contrast, the TUA could recommend this trade-off.

The democratic constraint of extreme measures may simply be a form of collective self-interest. Voters are unlikely to tolerate global catastrophic risks (GCRs), which incur the death of a sizeable portion of the electorate, if they know they themselves could be affected. We expect that scholars who do not support sacrificing current lives in the name of abstract calculations, but would still like to explore the use of expected value theory in existential risk, will be in support of democratic fail-safe mechanisms.

Empirically, this fail-safe mechanism seems to work. Even deeply imperfect democracies, like the ones we inhabit now, often avert detrimental outcomes. Democracies prevent famines[119] (although not malnutrition)[120]. They make war — a significant driver of GCRs — less likely[121]. The inclusion of diverse preferences in democracies, such as those achieved through women's suffrage, further decreases the likelihood of violent conflict[122]. Citizens often show a significant risk aversion in comparison to their government. While surveys are notoriously difficult to collect and interpret, existing data suggest that the public has little support for nuclear weapons use[123–125], but strong support for action against climate catastrophe[126–128]. We can further show that when citizens deliberately engage with the subject at hand, their concern and readiness for action often increases[118]. For example, citizen assemblies on climate change have recommended widespread policy-changes across sectors, amendments to incentive structures and laws against ecocide to reach emissions targets[129]. Indeed, many lament that when it comes to genetically modified organisms and nuclear power, citizens are far too risk-averse[130]. The problem is not that the public is riddled with cognitive biases that make them unconcerned about global catastrophes.

Democratic debate cannot be an afterthought. Navigating humanity through crises will involve many value-laden decisions under deep uncertainty. Democratic procedures can deal with such hard choices. Greater cognitive diversity should be represented amongst scholars of ERS. Recommendations on policies that would reduce risk should be passed through deliberative assemblies and await the approval of a wider pool of ordinary citizens, as they will be the ones who will bear this risk. A homogenous group of experts attempting to directly influence powerful decision-makers is not a fair or safe way of traversing the precipice.

## 7. Conclusion

The case for the importance of studying existential risk has been made. ERS must now converge on a trustworthy methodology.

The general problems in ERS we identify are that (i) an inclusive definition of existential or extinction risk will require some ambiguity, (ii) any categorisation of risks will be at least partly arbitrary, (iii) any risk assessment will not be entirely comprehensive, and (iv) any study of existential risks and proposed interventions could also increase risk. This field of study is inseparable from a moral inquiry. Definitions of what are catastrophic or desirable futures are inextricably intertwined with questions of value. Dealing with risk is not restricted to risk analysis but includes the question of what risks are worth taking. We believe these challenges are not insurmountable and are worthy of our attention.

The original and influential techno-utopia approach (the TUA) faces specific, daunting problems. These include idiosyncratic and non-representative moral visions of the future, and the use of definitions that are excessively ambiguous and founded on opaque, questionable assumptions. The categorisation problem is exacerbated by combining the study of catastrophe with longtermist ethics. The frameworks are crude and do not include recent advances in risk assessment, or even basic knowledge from directly relevant fields. The TUA does not yet consider response risks and at times advocates for risky policies. It is susceptible to being misused to justify exceptional emergency actions.

We suggest some initial, modest steps for improving the field. First, extinction risks should be analysed separately from extinction ethics and existential ethics. Some research questions will still require combining these inquiries, but the attempt to separate the science of risk from the moral evaluation of risk will benefit each endeavour. Second, existential risk scholars should transparently acknowledge the moral and empirical assumptions used in risk analyses. Third we must critically embrace the latest advances in risk assessment from other fields, such as climate science and disaster risk reduction. Fourth, and most importantly, existential risk must be cognitively diversified, and the judgement of its recommendations democratised. We can't afford to wait for a "Long Reflection". Open democracy and collective deliberation need to be central to reducing existential risk and navigating the future.

We encourage existing scholars to enrich the diversity of available frameworks by revising or abandoning the TUA. We encourage researchers to find entirely new approaches and take a more inclusive, participatory approach to thinking about and shaping our responses to potential catastrophes. To have good judgement, represent the interests of the vulnerable, and avoid dangers, the study of existential risk needs to be democratised.

## Acknowledgements

We would like to thank the almost two dozen reviewers who took the time to peruse this draft and provide invaluable suggestions. The paper is much stronger due to your help.